\documentclass[10pt,superscriptaddress,pra,twocolumn]{revtex4}%
\usepackage{amsfonts}
\usepackage{amsmath}
\usepackage{amssymb}
\usepackage{graphicx}
\usepackage{graphics}
\usepackage[usenames]{color}%
\begin{document}
\title{\textbf{Quantum-defect analysis of $3p$ and $3d$ \\ $\rm{H_3}$ Rydberg energy levels}}
\author{Jia Wang}
\affiliation{Department of Physics and JILA, University of Colorado,
Boulder, CO 80309}
\author{Chris H.~Greene}
\affiliation{Department of Physics and JILA, University of Colorado,
Boulder, CO 80309}

\begin{abstract}
In this work, Rydberg energy levels of the triatomic hydrogen
molecule ($\rm{H_3}$) are studied with multichannel quantum-defect
theory. We extract the body-frame p-wave quantum defects from highly
accurate \emph{ab initio} electronic potential surfaces and
calculate the quantum defects of higher angular momentum states in a
long-range multipole potential model. Laboratory-frame
quantum defect matrices emerge from a rovibrational-frame transformation carried out
with accurate rovibrational states of $\rm{H_3^+}$.  Finally, we use
the laboratory-frame quantum defects to calculate Rydberg energy
levels for the fundamental neutral triatomic molecule $\rm{H_3}$.

\end{abstract}
\maketitle

\section{Introduction}
The triatomic hydrogen molecule ($\rm{H_3}$) plays an important role
in astrophysics because its cation $\rm{H_3^+}$ acts as a
proton donor in chemical reactions occurring in interstellar clouds.
As the simplest triatomic neutral molecule, $\rm{H_3}$ also attracts
fundamental interest. Ever since its emission spectra were first
observed by G. Herzberg in the 1980s \cite{Herzberg1982}, $\rm{H_3}$
has been studied extensively. Herzberg and co-workers measured
infrared and visible emission spectra of $\rm{H_3}$ in discharges
through hydrogen and assigned them to Rydberg-Rydberg transitions
between $n=2$ and $n=3$ electronic states using empirical fits
\cite{Herzberg1982}. Helm and co-workers investigated the higher
Rydberg states and ionization potentials of $\rm{H_3}$ by analyzing
the photoabsorption spectrum \cite{Helm1986}. In 2003, building on
previous work of Schneider, Orel and Suzor-Weiner \cite{Orel2000},
it was shown \cite{Kokoouline2003, Kokoouline2001} that intermediate Rydberg states
of $\rm{H_3}$ play an important role in the dissociative
recombination (DR) process, $\rm{H_3^+} +\  e^- \rightarrow \rm{H_3}
\rightarrow \rm{H_2}+ \rm{H}\ or\  \rm{H}+\rm{H}+\rm{H}$. Prior to
the study of Ref. \cite{Kokoouline2003,Kokoouline2001}, the large
discrepancy between the DR rate determined by experiment and previous
theory had not been resolved.
Refs. \cite{Kokoouline2003, Kokoouline2001} found that Jahn-Teller
effects in $\rm{H_3}$ neglected in previous theoretical studies
couple the electronic and nuclear degrees of freedom and generate a
relatively high DR rate via intermediate p-wave Rydberg-state
pathways. A recent alternative formulation developed by Jungen and
Pratt provides supporting evidence for this interpretation
\cite{Jungen2009}. Vervloet and Watson improved both
the experimental techniques and empirical fits and reinvestigated the
low Rydberg states that G. Herzberg had observed \cite{Watson2003}.
Here we undertake an analysis of the Rydberg states with \emph{ab
initio} theory. One of the most successful techniques in treating
Rydberg states by \emph{ab initio} theory is multichannel quantum
defect theory (MQDT) \cite{Seaton1983, GreeneJungen1985}. Earlier
studies \cite{Kokoouline2003, Takagi1993, Schneider2008, Giusti1980} have utilized MQDT to
successfully describe the DR process.

The application of MQDT to study molecular Rydberg
energy levels treats the $\rm{H_3}$ molecule as a
Rydberg electron attached to a $\rm{H_3^+}$ ion. The
interaction between the Rydberg electron and the ion core is described through a smooth
reaction matrix $K$ or quantum defect matrix $\mu$. $K$ and $\mu$ are
simply related, e.g., for a single-channel,
$K=\tan\left(\pi\mu\right)$. We extract a body-frame reaction matrix
from \emph{ab initio} electronic potential surfaces for p-wave
Rydberg states and calculate the body-frame reaction matrix for
Higher angular momentum states ($l>1$) by using the long-range multipole
potential model. For the higher angular-momentum states, we neglect
short-range interactions due to the nonpenetrating nature of the
high $l$ states. Here $l$ denotes the quantum number of the Rydberg electron orbital
angular momentum. We then construct the total laboratory-frame
reaction matrix $K$ through a rovibrational frame transformation,
obtaining
\begin{equation}\label{rvtransgeneral}
K_{ii'}  = \sum\limits_{\alpha ,\alpha '} {\left\langle i \right|\left.
\alpha  \right\rangle } \tilde K_{\alpha \alpha '} \left\langle {\alpha '}
\right|\left. {i'} \right\rangle.
\end{equation}
Here, $K_{ii'}$ is the laboratory-frame reaction matrix element between
the laboratory-frame eigenchannels $\left| i \right\rangle$ and
$\left| i' \right\rangle$, and $\tilde K_{\alpha \alpha '}$ is the
body-frame reaction matrix element between body-frame eigenchannels
$\left| \alpha \right\rangle$ and $\left| \alpha' \right\rangle$.
The rovibrational frame transformation is specified by the unitary
transformation $U_{i\alpha }  = \left\langle i \right.\left| \alpha \right\rangle$.

The process of constructing the rovibrational transformation is
similar to that described in Ref. \cite{Kokoouline2003}and is based on the rovibrational wave functions of $\rm{H_3^+}$. To calculate
them, there are two important approximations adopted in
Ref. \cite{Kokoouline2003}, the rigid rotator approximation
and the adiabatic hyperspherical approximation. The nonadiabatic
coupling between different adiabatic hyperspherical channels was
included in later studies by using the slow variable discretization (SVD)
approach in Ref. \cite{Tolstikhin1996, Kokoouline2007}. In the present study,
we abandon the rigid rotator approximation and consider the
Coriolis interaction. In this way, we obtain very accurate
rovibrational energy levels and wave functions of $\rm{H_3^+}$
that allow us to construct the rovibrational transformation.

After the rovibrational transformation described by
Eq.(\ref{rvtransgeneral}) is carried out, we obtain the laboratory-frame $K$
matrix and calculate the eigenenergies $E$ of the $\rm{H_3}$ molecule
by solving the secular equation \cite{AymarGreene1996}
\begin{equation}
\det \left| {\tan \left( {\pi \nu } \right) + K} \right| = 0,
\end{equation}
where $\nu$ is a diagonal matrix with elements
$\nu _{ii}  = 1/\sqrt {2\left( {E_i^{rv}  - E} \right)}$.
Here $E_i^{rv}$ denotes the $i$th rovibrational energy level.

The remaining of the article is organized as follows. Section 2
describes the detailed calculation of rovibrational states of
$\rm{H_3^+}$ and shows how to use them to construct the rovibrational
transformation. Section 3 describes the calculation of the p-wave
energy levels of $\rm{H_3}$ using \emph{ab initio} quantum defects.
Section 4 discusses the long-range multipole potential
model for higher angular momentum Rydberg states, and Section 5 gives our
conclusions.

\section{Rovibrational states of $\rm{H_3^+}$}
In the cation $\rm{H_3^+}$, three protons interact with each other
under a potential surface. The potential surface was created by
Refs. \cite{Cencek1998} and is sub-micro-hartree accurate. With
this potential surface, we solve the three-body Schrodinger
equation in the adiabatic hyperspherical representation. We use
the SVD method to include the nonadiabatic coupling between different
adiabatic channels. We also include the Coriolis interaction that
couples rotational angular momentum with vibrational angular momentum.

After separation of the center-of-mass motion, the three-body
system can be described by six coordinates, the Euler angles
$\alpha$, $\beta$, and $\gamma$ and the three interparticle distances
$r_{12}$, $r_{23}$, and $r_{31}$. Alternatively, the interparticle
distances $r_{ij}$ can be expressed in terms of hyperspherical
coordinates, i.e., the hyperspherical radius $R$ with hyperspherical
angles $\theta$ and $\varphi$ \cite{Suno2002,WhittenSmith1968}, where
\begin{subequations}\label{hypercoord1}
\begin{equation}
r_{12}  = 3^{ - 1/4} R\left[ {1 + \sin \theta \sin \left( {\varphi - \pi /6} \right)} \right]^{1/2},
\end{equation}
\begin{equation}
r_{23}  = 3^{ - 1/4} R\left[ {1 + \sin \theta \sin \left( {\varphi  - 5\pi /6} \right)} \right]^{1/2},
\end{equation}
\begin{equation}
r_{31}  = 3^{ - 1/4} R\left[ {1 + \sin \theta \sin \left( {\varphi  + \pi /2} \right)} \right]^{1/2}.
\end{equation}
\end{subequations}
The hyperangle $\theta$ ranges from $0$ to $\pi/2$, while the
hyperangle $\varphi$ ranges from $0$ to $2\pi$. The hyperradius $R$
extends from $0$ to infinity. In hyperspherical coordinates, the
Sch\"odinger equation is given by
\begin{equation}\label{SchEq3body}
\left[ { - \frac{1}{{2\mu R^5 }}\frac{\partial }{{\partial R}}R^5
\frac{\partial }{{\partial R}} + \frac{{\Lambda ^2 }}{{2\mu R^2 }} +
V\left( {R,\theta ,\varphi } \right)} \right]\Psi _i  = E_i^{rv} \Psi _i.
\end{equation}
The Coriolis interaction is included in Eq.(\ref{SchEq3body}) through
the squared ``grand angular-momentum operator'' $\Lambda^2$ which
is given by
\begin{equation}
\frac{{\Lambda ^2 }}{{2\mu R^2 }} = T_\theta   + T_{\varphi C}  + T_r,
\end{equation}
where,
\begin{equation}
T_\theta   =  - \frac{2}{{\mu R^2 \sin 2\theta }}
\frac{\partial}{{\partial \theta }}\sin 2\theta \frac{\partial }{{\partial \theta }},
\end{equation}
\begin{equation}
T_{\varphi C}  = \frac{2}{{\mu R^2 \sin ^2 \theta }}\left( {i
\frac{\partial }{{\partial \varphi }} - \cos \frac{{J_z }}{2}} \right)^2,
\end{equation}
and
\begin{equation}
T_r  = \frac{{J_x^2 }}{{\mu R^2 \left( {1 - \sin \theta } \right)}} +
\frac{{J_y^2 }}{{\mu R^2 \left( {1 + \sin \theta } \right)}} +
\frac{{J_z^2 }}{{2\mu R^2 }}.
\end{equation}
The operators $\left( {J_x ,J_y ,J_z } \right)$ are the body-frame
components of the total angular momentum of the ion, and
$\mu=m/\sqrt{3}$ is the three-body reduced mass of the system,
where $m$ is the atomic hydrogen mass. Solving the Sch\"odinger
equation Eq.(\ref{SchEq3body}) directly should in principle provide accurate
rovibrational energy levels, but it would require extensive computational
time and memory to diagonalize the full Hamiltonian matrix.
Instead, we break the problem into two steps:  first solve the hyperangular Sch\"odinger equation in the adiabatic
representation, and then later include the nonadiabatic coupling through the SVD method.
\subsection{Adiabatic representation}
In the adiabatic representation, the adiabatic potentials and
channel functions are defined as solutions of the adiabatic
eigenvalue equations,
\begin{equation}\label{SchEqAdiabatic}
\left[ {\frac{{\Lambda ^2 }}{{2\mu R^2 }} + \frac{{15}}{{8\mu R^2}} +
V\left( {R,\theta ,\varphi } \right)} \right]\Phi _\nu
\left( {\Omega ;R} \right) = U_\nu  \left( R \right)\Phi _\nu \left( {\Omega ;R} \right),
\end{equation}
whose solutions depend parametrically on $R$. For each $R$,
the set of $\Phi _\nu \left( {\Omega ;R} \right)$ is orthogonal,
complete, and can be used as a basis to expand the whole wave
function, e.g.,
\begin{equation}\label{ExpandAdiabatic}
\Psi_i \left( {R,\Omega } \right) = \sum\limits_\nu  {R^{-5/2}
F_\nu^i  \left( R \right)\Phi _{\nu} \left( {\Omega ;R} \right)}.
\end{equation}
Here we use $\Omega$ to denote the Euler angles and the two
hyperspherical angles. To solve Eq.(\ref{SchEqAdiabatic})
numerically, we expand $\Phi _\nu \left( {\Omega ;R} \right)$
in a set of basis $\Phi _{jm_2 K^ +  }^{N^ +  m^ +  g_I }$,
such that
\begin{equation}
\Phi _\nu  \left( {\Omega ;R} \right) = \sum\limits_{jm_2 K^ +  } {a_{jm_2 K^ +  }^{\left( \nu  \right)} \left( R \right)\Phi _{jm_2 K^ +  }^{N^ +  m^ +  g_I } },
\end{equation}
where $j$, $m_2$, $N^+$, $K^+$ and $g_I$ are quantum numbers
labeling basis functions in different degrees of freedom.
$\Phi _{jm_2 K^ +  }^{N^ +  m^ +  g_I }$ satisfies the permutation
symmetry of a three-fermion system, e.g.,
\begin{subequations}
\begin{equation}\label{permusym1}
P_{12} \Phi _{jm_2 K^ +  }^{N^ +  m^ +  g_I }  =  -
\Phi _{jm_2 K^ +  }^{N^ +  m^ +  g_I },
\end{equation}
and
\begin{equation}\label{permusym2}
\mathcal{A} \Phi _{jm_2 K^ +  }^{N^ +  m^ +  g_I }  =
\Phi _{jm_2 K^ +  }^{N^ +  m^ +  g_I },
\end{equation}
\end{subequations}
where
\begin{equation}\label{permusym3}
\mathcal{A} = 1 - P_{12}  - P_{\,23}  - P_{31}  + P_{12} P_{31}  + P_{12} P_{23}.
\end{equation}
The explicit form of the basis functions $\Phi _{jm_2 K^ +  }^{N^ +
m^ +  g_I }$ is given in Appendix A.

\subsection{Slow variable discretization}
With the help of the discrete variable representation (DVR) basis
$\pi _n \left( R \right)$, effects of nonadiabatic couplings
between different adiabatic potentials are included. The DVR basis has the following properties:
\begin{equation}
\int {dR\pi _n \left( R \right)U_{\nu}\left( {R} \right)\pi _{n'}
\left( R \right)}  \approx U_{\nu}\left( {R_n} \right)\delta _{nn'},
\end{equation}
where $R_n$ are the quadrature abscissas used to generate the DVR
basis. The SVD approximation expresses the expansion of
Eq.(\ref{ExpandAdiabatic}) as
\begin{equation}
\Psi_i \left( {R,\Omega } \right) = \sum\limits_{\nu ,n}
{c_{n\nu}^i R^{-5/2} \pi _n \left( R \right)\Phi _\nu  \left( {\Omega ;R_n } \right)},
\end{equation}
and rewrites Eq.(\ref{SchEq3body}) as
\begin{equation}\label{ErvEq}
\sum\limits_{n,\mu } {T_{nn'} O_{n\nu ,n'\mu } c_{n'\mu }^i }
+ \left[ {U_{\nu} \left( {R_n } \right) - E^{rv}_i} \right]c_{n\nu}^i   = 0,
\end{equation}
where
\begin{equation}
T_{nn'}  = \int {dR\pi _n \left( R \right)\left[ { -
\frac{1}{{2\mu }}\frac{{\partial ^2 }}{{\partial R^2 }}}
\right]} \pi _{n'} \left( R \right)dR,
\end{equation}
and $O_{n\nu ,n'\mu }$ is the overlap matrix given by
\begin{equation}
O_{n\nu ,n'\mu }  = \left\langle {\Phi _\nu
\left( {\Omega ;R_n } \right)} \right|\left. {\Phi _\mu
\left( {\Omega ;R_n' } \right)} \right\rangle.
\end{equation}
Finally, Eq.(\ref{ErvEq}) is solved for the expansion coefficients
$c_{n\nu }^i$ and $E^{rv}_i$. The total rovibrational wave function
is therefore given by
\begin{equation}
\Psi _i \left( {R,\Omega } \right) = \sum\limits_{n\nu }
{c_{n\nu }^{i} \pi _n \left( R \right) \sum\limits_{jm_2 K^ +  }
{a_{jm_2 K^ + }^{\left( \nu \right)} \left( {R_n } \right)
\Phi _{jm_2 K^ + }^{N^ +  m^ + g_I } } },
\end{equation}
corresponding to the eigenenergy $E^{rv}_i$. Here $i$ is the
set of good quantum numbers $N^ +  m^ +  v^ +  g_I \Pi ^ +$,
where $v^ +$ denotes the vibrational quantum numbers.
\subsection{Accuracy of rovibrational energies of $\rm{H_3^+}$}
Next we
compare our theoretical rovibrational energy levels $E_i^{rv}$
of $\rm{H_3^+}$ with experimental energy levels
\cite{McCall2001}. Adopting the notation used in Ref.
\cite{McCall2001}, we label the rovibrational states $i$ by
quantum numbers $\left({N^+,G}\right)$$\{v_1,v_2^{l_2}\}(l|u)$.
$v_1$ is the symmetric-stretch vibrational quantum number, $v_2$
denotes the quantum number of the asymmetric-stretch mode, $l_2$
describes the quantum number of the vibrational angular momentum,
and $G \equiv \left| {K^ +   - l_2 } \right|$. The fact that
$G$ instead of $K^+$ is a good quantum number implies that the
Coriolis interaction couples rotational and vibrational angular
momenta and makes levels with the same $G$ nearly degenerate.
However, for levels with $l_2 \neq 0$ and $\left( {N^ +   - \left|
{l_2 } \right|} \right) \ge G \ge 1$, the degeneracy breaks,
and we $u$ (or $l$) to denote the upper (or lower)
energy level; these levels with a $u$ or an $l$
cannot be described by rigid rotator approximations.
\begin{table}[t]
\caption{Comparison of several calculated rovibrational energy levels of
$\rm{H_3^+}$ with experimental results \cite{McCall2001}. Only
states with $N^+ \leq 3$ for the $\{0,0^0\}$ and $\{0,1^1\}$ bands
are shown here.}
\begin{center}
\begin{tabular}{c|c|c||c|c|c}
\hline
Q.N.\footnotemark[1] & $E_{cal}$\footnotemark[2] & $E_{exp}$
\footnotemark[3] & Q.N.\footnotemark[1] & $E_{cal}$\footnotemark[2]
& $E_{exp}$ \footnotemark[3]\\
 & $\left(\rm{cm}^{-1}\right)$ & $\left(\rm{cm}^{-1}\right)$ &
 & $\left(\rm{cm}^{-1}\right)$ & $\left(\rm{cm}^{-1}\right)$\\
\hline
$(1,1)\{0,0^0\}$ & 64.128  & 64.121 & $(2,3)\{0,1^1\}$ & 2614.034 & 2614.270 \\
$(1,0)\{0,0^0\}$ & 86.960  & 86.960 & $(2,2)\{0,1^1\}$ & 2723.708 & 2723.962 \\
$(2,2)\{0,0^0\}$ & 169.288 & 169.295 & $(2,1)\{0,1^1\}l$ &  2755.313 &  2755.565 \\
$(2,1)\{0,0^0\}$ & 237.335 & 237.356 & $(2,1)\{0,1^1\}u$ &  2790.086 &  2790.344 \\
$(3,3)\{0,0^0\}$ & 315.317 & 315.349 & $(2,0)\{0,1^1\}$ & 2812.504 & 2812.850 \\
$(3,2)\{0,0^0\}$ & 427.974 & 428.018 & $(3,3)\{0,1^1\}$ & 2876.566 & 2876.847 \\
$(3,1)\{0,0^0\}$ & 494.712 & 494.775 & $(3,2)\{0,1^1\}l$ &  2931.091 &  2931.366 \\
$(3,0)\{0,0^0\}$ & 516.823 & 516.873 & $(3,2)\{0,1^1\}u$ &  2992.151 &  2992.436 \\
$(0,1)\{0,1^1\}$ & 2521.183 & 2521.411 & $(3,1)\{0,1^1\}l$ &  3002.348 &  3002.905 \\
$(1,2)\{0,1^1\}$ & 2547.996 &  2548.164 & $(3,0)\{0,1^1\}$ & 3025.663 & 3025.941 \\
$(1,1)\{0,1^1\}$ & 2609.302 & 2609.541 & $(3,1)\{0,1^1\}u$ &  3063.181 &  3063.951 \\
\hline
\hline
\end{tabular}
\end{center}
\begin{flushleft}
$^a$The quantum numbers labeling the energy levels, in the notation
$\left({N^+,G}\right)$$\{v_1,v_2^{l_2}\}(l|u)$ described in the text.\\
$^b$Theoretically calculated results from this work.\\
$^c$Experimentally determined energies from Ref. \cite{McCall2001}.\\
\end{flushleft}
\end{table}

Table $2$ compares the rovibrational energy levels, calculated for
$N^+ \leq 3$ states of $\{0,0^0\}$ and $\{0,1^1\}$ bands,  with the experimental results of Lindsay and
McCall.\cite{McCall2001} The agreement is good, with a  rms difference of $0.281$ $\rm{cm}^{-1}$ for the levels shown in the table.
Higher-rovibrational energy-level calculations also exhibit good
agreement. For our calculated energy levels up to around $9000$
$\rm{cm}^{-1}$ with $N^+ \leq 4$, the rms difference
between our calculation and the experimental results of
Ref. \cite{McCall2001} is $0.657$ $\rm{cm}^{-1}$.

\subsection{Rovibrational-frame transformation}
Next we describe in detail how to construct the rovibrational-frame
transformation using the ionic rovibrational eigenstates. In the laboratory frame,
the $H_3^+ + e^-$ system is described by the electron orbital
angular momentum $l$ and its projection $\lambda$ onto the
laboratory $z$-axis, and by $N^+$, $m^+$, $g_I$, and parity of the
ion core. Hence, we construct the wave function of the
$H_3^+ + e^-$ system as a sum of products of the ionic rovibrational
wave function and electronic wave function, of the form
\begin{equation}
\Psi_{N^ +  m^ +}^{v^ +  g_I \Pi ^ + } \left( {R,\Omega }
\right)Y_{l\lambda } \left( {\theta _e ,\varphi _e } \right),
\end{equation}
where $\theta _e$ and $\varphi _e$ are spherical angles of
the electron in the laboratory frame. The radial part of
electronic wave function is not shown here since we
apply the MQDT method for that degrees of freedom. Specifically, we want to construct
the laboratory eigenchannel function $\left| {i} \right\rangle$
with a definite laboratory-frame total angular momentum $N$: $\left| {i} \right\rangle = \left| {N^ + ,\nu ^ + } \right\rangle
^{\left(N,l,m,\Pi^+,g_I\right)}$. In the position representation, it takes the form as,
\begin{equation}\label{lfbasis}
\sum\limits_\lambda {C_{N^ +  ,m^ +  ;l,\lambda }^{N,m} \Psi _{N^ +  m^ +}^{v^ +  g_I \Pi ^ +  } Y_{l\lambda } },
\end{equation}
where $C_{N^ +  ,m^ +  ;l,\lambda }^{N,m}$ denotes the appropriate
Clebsch-Gordan coefficient.

In the body-frame, the is specified state by the projection of theelectron angular momentum on the molecular $Z$-axis $\Lambda$, by the total
angular momentum of the system $N$ including the electron contribution
$l$ and the projection $m$ of $N$ on the laboratory $z$-axis.
Applying the transformation between body-frame and lab-frame
\begin{equation}
Y_{l\lambda } \left( {\theta _e ,\varphi _e } \right) =
\sum\limits_\Lambda  {\left[{D_{\lambda \Lambda }^l \left( {\alpha ,\beta ,\gamma }
\right)}\right]^* Y_{l\Lambda } \left( {\theta _e' ,\varphi _e' } \right)},
\end{equation}
the expansion of the product of two Wigner functions,
\begin{equation}
D_{ m^ +  K^ +  }^{N^ +  } D_{\lambda \Lambda }^l  = \sum\limits_N
{D_{mK}^N C_{N^ +  ,K^ +  ;l.\Lambda }^{N,K} C_{N^ +  m^ +  ;l,\lambda }^{N,m} },
\end{equation}
we have the following equation,
\begin{eqnarray}\label{eqalge}
&& \sum\limits_\lambda  {C_{N^ +  ,m^ +  ;l,\lambda }^{N,m} \left[{D_{m^ +  K^ + }^{N^ +  }}\right]^*
Y_{l\lambda } \left( {\theta _e ,\varphi _e } \right)} \\ \nonumber = && \sum\limits_\Lambda
{\left(  -  \right)^{l - \Lambda } C_{l, - \Lambda ;N,K}^{N^ +  ,K^ +  }
\left[{D_{mK}^N}\right]^* Y_{l\Lambda } \left( {\theta _e ',\varphi _e '} \right)},
\end{eqnarray}
with some manipulation of algebra, where $\theta_e'$ and $\varphi_e'$
are the spherical angles of the electron in the body-frame. Using
Eq. (\ref{eqalge}), we derive the transformation between the body-frame
and laboratory-frame states as,
\begin{equation}
\left\langle {\alpha} \right.\left| {i} \right\rangle = \left\langle
{R,\Omega; \Lambda} \right.\left| {N^ +  ,\nu ^ + }
\right\rangle^{\left(N,l,\Pi ^ +,g_I \right)}  =
{\tilde \Psi^{N m g_I \Pi^+} _{N^+ v^ + \Lambda }  },
\end{equation}
where $\tilde \Psi^{N m g_I \Pi^+} _{N^+ v^ + \Lambda }$ is defined as,
\begin{eqnarray}\label{rovitrans}
\tilde \Psi^{N m g_I \Pi^+} _{N^+ v^ + \Lambda } = && \sum\limits_{n\nu }
{c_{n\nu }^{\left( {v^ +  } \right)} \pi _i \left( R \right)} \nonumber \\ && \times
\sum\limits_{jm_2 K^ +  } {a_{jm_2 K^ + }^{\left( v \right)}
\left( {R_n } \right)} \tilde \Phi _{jm_2 N^+ K^ +  \Lambda }^{Nmg_I }.
\end{eqnarray}
The explicit form of $\tilde \Phi _{jm_2 N^+ K^ +  \Lambda }^{Nmg_I }$ is given in Appendix A.
The rovibrational frame transformation can be accomplished as follows:
\begin{widetext}
\begin{equation}\label{rvtransform}
K_{N^ +  v^ +  ;N^ {+'}  v^ {+'}  }^{\left( {Nmg_I \Pi ^ +  } \right)} =
\sum\limits_{\Lambda ,\Lambda '} {\int {d\mathcal{Q }} } d\Omega _E
\tilde \Psi _{N^ +  v^ +  \Lambda }^{Nm\Pi ^ +  g_I *} K_{\Lambda \Lambda' }
\left( {\mathcal{Q }} \right)\tilde \Psi _{N^ {+'}   v^ {+'}
\Lambda '}^{Nm\Pi ^ +  g_I },
\end{equation}
\end{widetext}
where $\Omega_E$ denotes the Euler angles, and $\mathcal{Q}$
contains the three vibrational degrees of freedom.
\section{p-wave energy levels of $\rm{H_3}$}
The body-frame reaction matrix for a p-wave electron is described by
the short-range interaction extracted from an \emph{ab initio}
calculation \cite{MJungen2000}. In practice, the quantum defects are
smoother than the reaction matrix elements because the latter can have poles. Hence in this work, we
extract the body-frame quantum defects $\mu \left( \mathcal{Q}
\right)$ from the \emph{ab initio} energy surface directly.
After replacing $K_{\Lambda \Lambda' } \left( {\mathcal{Q} } \right)$ by
$\mu_{\Lambda \Lambda' } \left( \mathcal{Q} \right)$ in
Eq.(\ref{rvtransform}), we perform a rovibrational transformation to
get the laboratory-frame quantum-defect matrix. Finally, we get the
laboratory $K$ matrix by using the eigenvalues $\mu_e$ from the
laboratory-frame quantum-defect matrix,
\begin{equation}
K = U\tan \left( {\pi \mu _e } \right)U^T,
\end{equation}
where $U$ denotes the unitary transformation that diagonalizes the
laboratory-frame quantum-defect matrix.
\subsection{Body-frame quantum defects for p-waves}
Because of Jahn-Teller effects, the body-frame quantum-defect matrix
is generally not diagonal in the electronic projections $\Lambda$,
$\Lambda'$. Similar to the body-frame $K$ matrix proposed by Staib
and Domcke \cite{Domecke1990}, it has the form
\begin{equation}\label{mumatrix}
\mu \left( \mathcal{Q} \right) = \left[ {\begin{array}{*{20}c}
{\mu _{00} \left( \mathcal{Q} \right)} & 0 & 0  \\
0 & {\mu _{11} \left( \mathcal{Q} \right)} & {\mu _{1 - 1} \left( \mathcal{Q} \right)}  \\
0 & {\mu _{ - 11} \left( \mathcal{Q} \right)} & {\mu _{ - 1 - 1} \left( \mathcal{Q} \right)}  \\
\end{array}} \right].
\end{equation}
We express each matrix element by using the vibrational symmetry
coordinates $\mathcal{Q}=\left({{Q}_1,{Q}_x,{Q}_y}\right)$ as
\begin{subequations}\label{hypercoord2}
\begin{equation}
Q_1  = f\frac{1}{{\sqrt 3 }}\left( {\Delta r_1  + \Delta r_2 + \Delta r_3 } \right),
\end{equation}
\begin{equation}
Q_x = f\frac{1}{{\sqrt 3 }}\left( {2\Delta r_3  - \Delta r_2  - \Delta r_1 } \right),
\end{equation}
and
\begin{equation}
Q_y  = f\left( {\Delta r_1  - \Delta r_2 } \right).
\end{equation}
\end{subequations}
where $f=2.639\ 255$ bohr$^{-1}$ is a constant and $\Delta r_i$ describe displacements of the nuclei from the
equilibrium geometry at which $r_{12}  = r_{23}  = r_{31}
= r_{equi} = 1.6504$ $a.u.$. For example, $\Delta r_1  = r_{23}  -
r_{\rm{equi}}$. $\left({{Q}_x,{Q}_y}\right)$ can be alternatively
described by another pair of coordinates $\left( {\rho ,\phi } \right)$
as $Q_x = \rho \cos \phi$ and $Q_y  = \rho \sin \phi$. $Q_1$
describes the symmetric stretch of the molecule, while
$\left({{Q}_x,{Q}_y}\right)$ or $\left( {\rho ,\phi } \right)$ describe
bends and the asymmetric stretch. These coordinates are very useful
here for our Taylor expansion of the body-frame quantum defects around an
equilibrium position. We use the following forms,
\begin{equation}\label{mufit1}
\mu _{00} \left( \mathcal{Q} \right) = \mu _{00} \left(
{\mathcal{Q} = 0} \right) + a_1 Q_1  + a_2 Q_1^2  +
a_3 Q_1^3 + a_4 \rho ^2  ,
\end{equation}
\begin{eqnarray}
\mu _{11} \left( \mathcal{Q} \right) && = \mu _{-1-1} \left( \mathcal{Q}
\right)  \\ && =\mu _{11} \left( {\mathcal{Q} = 0} \right) + b_1 Q_1  +
b_2 Q_1^2 + b_3 Q_1^3 + \delta  \rho ^2, \nonumber
\end{eqnarray}
and
\begin{equation}\label{mufit2}
\mu _{1 - 1} \left( \mathcal{Q} \right) = \mu _{ - 11} \left(
\mathcal{Q} \right) = {\lambda  }\rho.
\end{equation}
The form of our off-diagonal matrix elements $\mu _{1 - 1} \left(
\mathcal{Q} \right)$ and $\mu _{ - 11} \left( \mathcal{Q} \right)$
differs from the usual adopted form in Ref. \cite{Kokoouline2003} by
a phase factor $\exp{\left(\pm i\phi\right)}$. This different phase
convention is due to the fact that the usual adopted form was
derived in a body frame that is rotated from our body frame by $\phi/2$. We
develop the detailed proof in Apendix B. The effective quantum numbers
are calculated by diagonalizing the quantum defect matrix. This
calculation gives,
\begin{equation}
\nu _{n,\pi _{1,2} } \left( \mathcal{Q} \right) = n -  \left[ {\mu
_{11} \left( \mathcal{Q} \right) \pm \left| {\mu _{1 - 1} \left(
\mathcal{Q} \right)} \right|} \right],
\end{equation}
an expression that can be used to fit the effective quantum numbers calculated
\emph{ab-initio} in Ref. \cite{MJungen2000}.
\begin{table}[h]
\caption{A comparison between several of our calculated $3p_1$ $\rm{H_3}$
energy levels with empirically fitted experimental
energy levels\cite{Herzberg1982}.}
\begin{center}
\begin{tabular}{c|c|c|c}
\hline
Label\footnotemark[1] & $E_{cal}$\footnotemark[2] &
$E_{fit}$ \footnotemark[3] & Differences \footnotemark[4]\\
$N,g,U$ & $\left(\rm{cm}^{-1}\right)$ &
$\left(\rm{cm}^{-1}\right)$ & $\left(\rm{cm}^{-1}\right)$\\
\hline
$0,1,1$ & 12967.8  & 12966.863  &0.9\\
$1,0,1$ & 12999.1  & 12998.196  &0.9\\
$1,1,1$ & 13052.3 & 13050.966  &1.3\\
$1,2,1$ & 13066.9 & 13068.700  &-1.8\\
$2,0,-1$ & 13139.9 & 13138.608  &1.3\\
$2,1,-1$ & 13056.1 & 13056.588  &-0.5\\
$2,1,1$ & 13221.1 & 13219.125  &1.9\\
$2,2,1$ & 13234.3 & 13235.522  &-1.2\\
$2,-3,1$ & 13203.8 & 13212.055  &-8.3\\
$3,0,1$ & 13450.7 &  13446.072  &4.6\\
$3,1,-1$ & 13300.6 & 13300.119  &0.5\\
$3,2,-1$ & 13160.5 & 13165.030  &-4.5\\
$3,2,1$ & 13485.4 & 13483.545  &1.9\\
$3,3,1$ & 13453.2 & 13460.934 &-7.7 \\
\hline
\hline
\end{tabular}
\end{center}
\begin{flushleft}
$^a$The label denotes the values of $N,G,U$ adopted in Ref. \cite{Watson2003}
to fit the experimental energy levels.\\
$^b$Theoretical results calculated in this study.\\
$^c$Empirical fits for experimental energies determined in Ref. \cite{Watson2003}.\\
$^d$Differences theory - experiment.\\
\end{flushleft}
\end{table}
\subsection{$3p_1$ energy levels of $\rm{H_3}$}
We calculate the $3p_1$ energy levels of $H_3$ and compare them with
empirical fits from Ref. \cite{Watson2003}. Quantum defect
parameters in Eqs.(\ref{mufit1}--\ref{mufit2}) are extracted from
the \emph{ab-initio} calculation in Ref. \cite{MJungen2000}. To fit
the experiment results, we shift the quantum defects at equilibrium
positions $\mu _{00} \left( {\mathcal{Q} = 0} \right)= 0.0683$ and
$\mu _{11} \left( {\mathcal{Q} = 0} \right)=0.4069$ by a small
amount, $0.0043$ and $0.0021$ correspondingly.

In Ref. {\cite{Watson2003}}, Vervloet and Watson
studied the $H_3$ emmision lines of $\left( {3s,3p_0 ,3d} \right)
\to 2p_0$ bands and $\left( {3s,3p_0 ,3d} \right) \to 3p_0$ bands.
They then fitted the lines with effective Hamiltonians of the following form,
\begin{equation}
\begin{array}{l}
 BN\left( {N + 1} \right) + \left( {C - B} \right)K^2  \\
  - D_N N^2 \left( {N + 1} \right)^2  \\
  - D_{NK} N\left( {N + 1} \right)K^2  \\
  - D_K K^4  + \ldots \\
 \end{array}
\end{equation}
where the explicit expressions can be found in Ref. \cite{Herzberg1982}
and Ref. \cite{Watson2003}. Table $2$ compares our MQDT
result with the experimental energy levels calculated from the fitted effective
Hamiltonians. The labels $N,g,U$ are fitting parameters, where
$N$ is also the total angular momentum of $\rm{H_3}$, and $g$ is
related to the quantum number $G$ by $G=\left|g\right|$. Evidently our calculations are in good agreement with the fitted and recalculated
experimental results, with differences of around a few
$\rm{cm}^{-1}$.
\section{Higher angular-momentum states}
For higher electronic angular-momentum states with $l>1$,
the orbits are nonpenetrating and the short-range interaction is
negligible. The long-range multipole potential model employing
perturbation theory has successfully described the high orbital
angular Rydberg states of $\rm{H_2}$ \cite{Eyler1983, Jungen1982}.
In this work, we include the perturbations and interactions between levels of
different $n$ (principle quantum number) and $l$ (angular-momentum
quantum number) in a systematic fashion by incorporating the
formalism of MQDT \cite{Seaton1999}.  We use this long-range model
to calculate the Rydberg states of $\rm{H_3}$ with $l\geq2$.

For a Rydberg electron with high orbital angular momentum  ($l\geq2$
in the case of $\rm{H_3^+}$), the effects of core penetration are
negligible. Hence, the interaction between the Rydberg electron and
the ion core can be approximately described by two effects. First,
the interaction potential between the Rydberg electron and the
molecular ion is expanded into a multipole series, where the
quadrupole moment of the $\rm{H_3^+}$ core is the leading
anisotropic term. Second, the induced dipole moment of the ion core
interacts with the Rydberg electron by a potential characterized by
the polarizability of the $\rm{H_3^+}$ core. All higher angular
momenta and higher-order polarizabilities are neglected here, as well as the anisotropic portion of the polarizability interaction.
\begin{table}[t]
\caption{Comparison between several of our calculated $3d$ energy
levels of $\rm{H_3}$ with experimentally-determined energy
levels\cite{Herzberg1982}.}
\begin{center}
\begin{tabular}{c|c|c|c}
\hline
Label\footnotemark[1] & $E_{cal}$\footnotemark[2] &
$E_{fit}$ \footnotemark[3] & Differences\footnotemark[4]\\
$N^+,K^+,N$ & $\left(\rm{cm}^{-1}\right)$ &
$\left(\rm{cm}^{-1}\right)$ & $\left(\rm{cm}^{-1}\right)$ \\
\hline
$2,1,0$ & 17399.14 & 17415.86 & -15.89 \\
$2,2,0$ & 17058.41 & 17039.61 & 18.80 \\
$1,0,1$ & 17284.81 & 17296.57 & -11.76 \\
$3,0,1$ & 17742.32 & 17741.29 & 1.03 \\
$1,1,1$ & 17005.99 & 16991.72 & 14.28 \\
$2,1,1$ & 17403.89 & 17412.83 & -8.94\\
$3,1,1$ & 17698.40 & 17700.43 & -2.02  \\
$2,2,1$ & 17107.24 & 17094.12 & 13.13 \\
$3,2,1$ & 17540.96 & 17557.32 & -16.35 \\
$3,3,1$ & 17204.46 & 17188.48 & 15.98  \\
$1,0,2$ & 17011.36 & 17001.08 & 10.27 \\
$3,0,2$ & 17643.36 & 17655.58 & -12.21\\
\hline
\hline
\end{tabular}
\end{center}
\begin{flushleft}
$^a$The label denotes the quantum numbers  $N^+,K^+,N$.\\
$^b$Theoretical results calculated in this work.\\
$^c$Empirical fits to the experimental energies determined in Ref. \cite{Watson2003}.\\
$^d$Differences between theoretical and experimental results.\\
\end{flushleft}
\end{table}

In this approximation, the Hamiltonian is given in atomic units by
\begin{equation}
H =  - \frac{1}{2}\nabla ^2  - \frac{1}{r} + V_{eff}+H_{core},
\end{equation}
where $H_{core}$ is the rovibrational energy of the $H^+_3$ core.
$V_{eff}$ includes quadrupole and polarizability interactions:
\begin{equation}
V_{eff}  =  V_{quad}+V_{pol} = - \frac{{Q_2 }} {{r^3 }}P_2
\left( {\cos \theta_e ' } \right) - \frac{{\alpha}} {{2r^4 }} -
\frac{\gamma } {3r^4}P_2 \left( {\cos \theta_e ' } \right).
\end{equation}
where $Q_2$, $\alpha$, and $\gamma$ are respectively the quadrupole moment,
isotropic polarizability and the cylindrically-symmetric anisotropic polarizability. Other components of the quadrupole moment tensor
vanish for the undistorted equilateral triangle configuration. For the vibrational ground state, $Q_2$, $\alpha$, and $\gamma$ are
taken from table III of Ref. \cite{Dykstra1998}. The polarizability
and quadrupole interactions are much smaller than the Coulomb
interaction and hence will be treated in perturbation theory. We
also find that the quantum defect is small (of the order of 0.01),
and the coupling between vibrational ground states of $\rm{H_3^+}$
to excited vibrational states are negligible. Hence, in the
rovibrational transformation, we only include the vibrational ground
state. The body-frame reaction matrix thus can be written as
\begin{equation}
K_{\Lambda \Lambda '}  \approx  - \pi \int {drf_{nl}\left( r
\right)\left\langle {Y_{l\Lambda } } \right|V_{eff}\left|
{Y_{l\Lambda '} } \right\rangle f_{nl}\left( r \right)},
\end{equation}
where $f_{nl}$ is the regular Coulomb function with $l=1$ as the
angular momentum quantum number and $n$ as the principal quantum
number. As the quantum defect for $d$-wave electrons are small, we
can use integers for $n$ in calculating the radial functions. Here, $r$ is the electronic radial
coordinate. Again, we perform a rovibrational transformation (with
only the vibrational ground states) to obtain the laboratory-frame $K$
matrix and finally, calculate the energy levels using the standard determinantal equation of MQDT.

Table $3$ compares our theoretical calculations with the
experimentally-determined 3d energy levels \cite{Watson2003}. The
agreement somewhat poorer than the p-wave case.

\section{Summary}
In this work, we have calculated the Rydberg energy levels of
$\rm{H_3}$ molecules. Using an accurate \emph{ab-initio}
quantum-defect surface and \emph{ab-initio} core energies of
$\rm{H_3^+}$, our theoretical results for the p-wave Rydberg states
from the present MQDT calculations are in good agreement with
experimental results from J. K. G. Watson \cite{Watson2003}. We also
study higher-momentum states by a using a long-range multipole
potential model in conjunction with MQDT, and find encouraging agreement with
experimental results from Ref. \cite{Watson2003}.

\section*{Acknowledgements}
This work was supported in part by the Department of Energy, Office
of Science.  We thank V. Kokoouline and R. Saykally for helpful
discussions.

\appendix

\section{Permutation symmetry of the basis functions}
For convenience, we first consider trial basis functions as
\begin{equation}\label{trialbasis}
\Phi _{\rm{try}}  = e^{im_2 \varphi } \mathcal{R}_{K^ +  m^ +  }^{N^ +  }\left({\alpha,\beta,\gamma}\right)
\Phi _{g_I }^I u_j \left( \theta  \right),
\end{equation}
and use them to construct basis functions with proper permutation
symmetry Eqs.(\ref{permusym1}--\ref{permusym2}). The continuity
condition for Smith-Whitten hyperspherical coordinates
\cite{Kendrick1999,Suno2002} requires that,
\begin{equation}
\Phi _{\rm{try}} \left( {\theta ,\varphi ,\alpha ,\beta ,\gamma }
\right) = \Phi _{\rm{try}} \left( {\theta ,\varphi  + 2\pi ,\alpha
,\beta ,\gamma  + \pi } \right).
\end{equation}
This boundary condition leads to the condition that $K^+ / 2 + m_2$
must be integral. Hence, if $K^+$ is even, $m_2$ must be integral;
if $K^+$ is odd, $m_2$ must be half integral. We remark here that
the parity $\Pi^+$ is determined by $K^+$ only; $\Pi^+ = +1$, if
$K^+$ is even, and $\Pi^+=-1$, if $K^+$ is odd
\cite{Kendrick1999,Suno2002}.

The permutation symmetries for the basis functions chosen for each
degree of freedom are shown in Table $4$. Here, $\Phi^I_{g_I}$ is
the nuclear-spin basis function defined as in Ref.
\cite{Kokoouline2003}. $g_I$ equals zero for the ortho state, and
$\pm 1$ for the para state. The rotational part $\mathcal{R}_{K^ +  m^ +  }^{N^ +  }\left({\alpha,\beta,\gamma}\right)$
is given by,
\begin{equation}
\mathcal{R}_{K^ +  m^ +  }^{N^ +  }\left({\alpha,\beta,\gamma}\right)=
\sqrt{\frac{{2N^+ + 1}}{{8\pi ^2 }}} \left[{D^{N^ +}_{m^+ K^+ }\left({\alpha,\beta,\gamma}\right)}\right]^*
\end{equation}
where $D^{N^ +}_{m^+ K^+}$ are the Wigner D functions of the Euler angles. The phase of the Wigner function is chosen as in
Varshalovich \emph{et al}. \cite{Varshalovich}. $N^+$ is the total angular momentum
of the ion, $K^+$ is the projection of $N^+$ onto the body
frame z-axis, and $m^+$ is the projection onto the laboratory frame
Z-axis. We also use a set of fifth-order basis splines
$u_j\left({\theta}\right)$ to expand the wave function in $\theta$. Since $u_j\left({\theta}\right)$ is unaffected by
permutations, it is not shown in Table 4.
\begin{table}[t]
\caption{Permutation symmetry for basis functions in the different
degrees of freedom.}
\begin{center}
\begin{tabular}{c|c|c|c}
\hline
Permutation & $e^{im_2\varphi}$ & $\mathcal{R}^{N^ +}_{K^+ m^+}$ & $\Phi^I_{g_I}$\\
Operation& & &\\
\hline
$P_{12}$ & $e^{i4\pi/3}e^{-im_2\varphi}$ & $\left({-}
\right)^{N^+ + K^+}\mathcal{R}^{N^ +}_{-K^+ m^+}$ & $e^{i4\pi g_I/3}\Phi^I_{-g_I}$\\
$P_{23}$ & $e^{i2\pi/3}e^{-im_2\varphi}$ & $\left({-}
\right)^{N^+}\mathcal{R}^{N^ +}_{-K^+ m^+}$ & $e^{i2\pi g_I/3}\Phi^I_{-g_I}$\\
$P_{31}$ & $e^{i2\pi}e^{-im_2\varphi}$ & $\left({-}
\right)^{N^+ + K^+}\mathcal{R}^{N^ +}_{-K^+ m^+}$ & $e^{i2\pi g_I}\Phi^I_{-g_I}$\\
$P_{12}P_{31}$ & $e^{i2\pi/3}e^{im_2\varphi}$ & $\left({-}
\right)^{K^+}\mathcal{R}^{N^ +}_{K^+ m^+}$ & $e^{i2\pi g_I/3}\Phi^I_{g_I}$\\
$P_{12}P_{23}$ & $e^{i4\pi/3}e^{im_2\varphi}$ & $\mathcal{R}^{N^ +}_{K^+ m^+}$ & $e^{i4\pi/3g_I}\Phi^I_{-g_I}$\\
\hline
\end{tabular}
\end{center}
\end{table}

Application of the antisymmetrization operator $\mathcal{A}$ in
Eq.(\ref{permusym3}) to Eq.(\ref{trialbasis}) leads to:
\begin{eqnarray}\label{antibasis}
\mathcal{A}\Phi _{\rm{try}}  = && u_j \left( \theta  \right) \left[ {1 + \left(  -  \right)^{K^{+}} e^{i\frac{{2\pi }}{3}
\left( {m_2 + g_I } \right)}  + e^{i\frac{{4\pi }}{3}
\left( {m_2 + g_I } \right)} } \right] \nonumber \\ && \times \left[ {e^{im\varphi }
\mathcal{R}_{K^ +  m^ +  }^{N^ +  } \Phi _{g_I }^I  - e^{i\frac{{2\pi }}{3}
\left( {m_2 + g_I } \right)} \left(  -  \right)^{N^ +  }}\right. \nonumber \\ && \times
\left. {e^{ - im\varphi } \mathcal{R}_{ - K^ +  m^ +  }^{N^ +  } \Phi _{ - g_I }^I } \right]
\end{eqnarray}
Following the fact that $m_2$ is integral (half integral) if $K^+$
is even (odd), and that $g_I$ equals $0$ or $\pm 1$, we see that the
factors in the second line of the right hand side of
Eq.(\ref{antibasis}) vanish unless,
\begin{subequations}
\begin{equation}\label{qncondition1}
m_2  + g_I  = 3n \ \rm{if} \ K^+ \ \rm{is even},
\end{equation}
\begin{equation}\label{qncondition2}
m_2  + g_I  = 3n+3/2 \ \rm{if} \ K^+ \ \rm{is odd},
\end{equation}
\end{subequations}
where $n$ is any integer. Under the conditions
Eqs.(\ref{qncondition1}--\ref{qncondition2}), the factors in the
first line of the right hand side of Eq.(\ref{antibasis}) vanish
if $m_2=0$, $g_I=0$, $K^+=0$ when $N^+$ is even. Therefore, the basis functions that obey the permutation symmetry
are:
\begin{equation}
\Phi _{jm_2 K^ +  }^{N^ +  m^ +  g_I }  = u_j\left({\theta}\right) e^{i m_2 \varphi }
\mathcal{R}_{K^ + m^ +  }^{N^ +  } \Phi _{g_I
}^I ,
\end{equation}
if $m_2=0$, $K^+=0$, $g_I=0$, and $N^+$ is odd, otherwise,
\begin{eqnarray}
\Phi _{jm_2 K^ +  }^{N^ +  m^ +  g_I } && =  \frac{1}{{\sqrt 2 }}
u_j\left({\theta}\right) \left[ {e^{im_2 \varphi } \mathcal{R}_{K^ + m^ +
}^{N^ +  } \Phi _{g_I }^I}\right. \nonumber \\   &&\left.{- \left(  -  \right)^{N^ + + K^ +} e^{ -
im_2 \varphi } \mathcal{R}_{-K^ +  m^ +  }^{N^ +  } \Phi _{ - g_I }^I }
\right],
\end{eqnarray}
where $m_2$, $g_I$ and $K^+$ obey the relations Eqs. (\ref{qncondition1}--\ref{qncondition2}).

The explicit form of $\tilde \Phi _{jm_2 N^+ K^ +  \Lambda }^{Nmg_I }$ in Eq. (\ref{rovitrans}) is
closely related to $\Phi _{jm_2 K^ +  }^{N^ +  m^ +  g_I }$. Insertion of Eq.(\ref{eqalge}) into Eq.(\ref{lfbasis}) yields
\begin{eqnarray}
\tilde \Phi _{jm_2 N^+ K^ +  \Lambda }^{Nmg_I }  = && u_j \left( \theta
\right)\left(  -  \right)^{l - \Lambda } e^{im_2 \varphi } \nonumber \\
&&\times \Phi _{g_I }^I C_{l, - \Lambda ;N,K}^{N^ +  K^ +  }
\mathcal{R}_{Km}^N ,
\end{eqnarray}
with $K=K^+ + \Lambda$, if $m_2=0$, $g_I=0$, $K^+=0$ and $N^+$ is odd. Otherwise,
\begin{eqnarray}
\tilde \Phi _{jm_2 N^+ K^ +  \Lambda }^{Nmg_I }  && =  \frac{{u_j
\left( \theta  \right)}}{{\sqrt 2 }}\left(  -  \right)^{l - \Lambda } \nonumber \\ && \times \left[ {e^{im_2 \varphi }
\Phi _{g_I }^I C_{l, - \Lambda ;N,K}^{N^ +  K^ +  } \mathcal{R}_{Km}^N -
\left(  -  \right)^{N^ +   + K^ + } }\right. \nonumber \\ && \left.{  \times e^{ - im_2 \varphi }
\Phi _{ - g_I }^I C_{l, - \Lambda ;N,\tilde K}^{N^ +   - K^ +  }
\mathcal{R}_{\tilde Km}^N } \right],
\end{eqnarray}
where $\tilde K=-K^+ + \Lambda$.

\section{Body-frame quantum defect matrix elements}
The form of reaction matrix $K$ describing the Jahn-Teller
coupling of the p-wave electron can be written as
\begin{equation}
K_{\Lambda \Lambda '}  = \left( {\begin{array}{*{20}c}
   {K_{00} } & {K_{01}} & {K_{0-1}} \\
   {K_{10}} & {K_{11} } & {K_{1 - 1} }  \\
   {K_{-10}} & {K_{ - 11} } & {K_{ - 1 - 1} }  \\
\end{array}} \right).
\end{equation}
In perturbation theory, the matrix elements of the reaction matrix near the equilibrium position obey,
\begin{eqnarray}\label{Klambda}
K_{\Lambda \Lambda '}  \approx  && - \pi \int {dr'{\bar {f}_{\varepsilon l \Lambda} \left({r'}\right)\bar {f}_{\varepsilon l \Lambda'}\left({r'}\right) }}\\
&& \times \int {d\omega _e'Y_{l\Lambda }^* \left( {\theta _e' ,\varphi _e'} \right)} \Delta
V_e \left( {\mathcal{Q}; \mathbf{r_e}'} \right) Y_{l\Lambda '} \left( {\theta _e' ,
\varphi _e' } \right), \nonumber
\end{eqnarray}
where $\Delta V_e  = V_e \left( {\mathcal{Q}; \mathbf{r_e}'} \right) - V_e \left( {\mathcal{Q}=0; \mathbf{r_e}'} \right)$, and $V_e \left( {Q,r_e '} \right)$
is the interaction between the ion and the Rydberg
electron excluding the Coulomb potential, $\mathcal{Q}$ are the vibrational symmetry coordinates, and $\mathbf{r_e}'=
\left\{ {r ',\theta _e ',\varphi _e '} \right\}$ are the spherical coordinates of the electron in the body-frame.
$\bar {f}_{\varepsilon l \Lambda} \left({r'}\right)$ is the regular phase-renormalized Coulomb wave function defined by,
$\bar f_{\varepsilon l\Lambda }  = f_{\varepsilon l} \cos \left( {\pi \mu _{\Lambda \Lambda } \left( {\mathcal{Q} = 0} \right)} \right) - g_{\varepsilon l} \sin \left( {\pi \mu _{\Lambda \Lambda } \left( {\mathcal{Q} = 0} \right)} \right)$, where $\left\{ {f_{\varepsilon l},g_{\varepsilon l}} \right\}$ are the usual Coulomb wave functions with energy $\varepsilon$ and angular momentum $l$. $\mu _{\Lambda \Lambda } \left( {\mathcal{Q} = 0} \right)$ are the constant zero-order term of the diagonal quantum defect matrix elements.

To explore the symmetry properties of $K_{\Lambda \Lambda '}$, we write the Taylor expansion
in terms of $\mathcal{Q}=
\left\{ {Q_1,Q_+,Q_-} \right\}$, where $Q_ +   = \rho e^{+ i\phi }$ and $Q_ -   =
\rho e^{- i\phi }$,
\begin{equation}\label{expandV}
K_{\Lambda \Lambda '}  =
\sum\limits_\nu   {K_{\Lambda \Lambda '}^\nu  Q_\nu  }  +
\frac{1}{2}\sum\limits_{\nu \mu } {K_{\Lambda \Lambda '}^{\nu \mu }
Q_\nu  Q_\mu  }  + \ldots
\end{equation}
where each summation is over the subscripts $1$, $+$ and $-$.
However,  the expansion Eq. (\ref{expandV}) is not valid in our body
frame coordinates because $\Delta V_e \left( {\mathcal{Q}; \mathbf{r_e}'} \right)$ is not single-valued at $Q_+=Q_-=0$. This fact is demonstrated in the following discussion of our body frame
coordinates.

The body frame coordinates used here are the same as the coordinates
defined in Eq. (7) of Ref. \cite{Suno2002}, but we denote them as
$x'y'z'$ instead of $xyz$ to make our notations consistent. The $z'$-axis
is perpendicular to the plane defined by the three nuclei and the
$x'$-axis is associated with the smallest moment of inertia.
After manipulating Eqs.(2), (3), and Eq.(7) of Ref. \cite{Suno2002} with
some algebra, we can write down the cartesian coordinates of the
positions of the $i$th nuclei $\left( {x_i ',y_i', z_i'} \right)$
as
\begin{subequations}
\begin{equation}
x_i ' = \frac{2}{{3d}}R\cos \left( {\frac{\theta }{2} - \frac{\pi }{4}}
\right)\cos \left( {\frac{\varphi }{2} + \vartheta _i } \right),
\end{equation}
\begin{equation}
y_i ' =  - \frac{2}{{3d}}R\sin \left( {\frac{\theta }{2} - \frac{\pi }{4}}
\right)\sin \left( {\frac{\varphi }{2} + \vartheta _i } \right),
\end{equation}
\begin{equation}
z_i ' = 0.
\end{equation}
\end{subequations}
where $\vartheta _1 = 5\pi/6$, $\vartheta _2 = - \pi/2$ and $\vartheta _3 =
\pi/6$. When $\theta$ is very small, following Eqs. (\ref{hypercoord1}) and Eqs. ({\ref{hypercoord2}})
the two sets of coordinates $\left\{ {Q_1, \rho, \phi} \right\}$
and $\left\{ {R, \theta, \varphi } \right\}$ have the following relationship,
\begin{subequations}
\begin{equation}
Q_1  = 3^{1/4} f\left( {R - R_0 } \right),
\end{equation}
\begin{equation}
\rho  = 3^{1/4} fR \theta /2,
\end{equation}
\begin{equation}
\phi  = \varphi  - 2\pi /3.
\end{equation}
\end{subequations}
where $R_0=3^{1/4}r_{\rm{equi}}$. Therefore, when $Q_+=Q_-=0$, and
hence $\rho=0$, the positions of the $i$th nuclei $\left( {x_i ',y_i', z_i'}
\right)$ can be written as,
\begin{subequations}
\begin{eqnarray}
&& x_i '\left( {Q_1 ,Q_ +   = 0,Q_ -   = 0} \right) = \nonumber \\
&& \frac{{\sqrt 2 }}{{3d}}\left( {R_0  + \frac{{Q_1 }}{{3^{1/4} f}}} \right)
\cos \left( {\frac{\phi }{2} + \vartheta _i  + \frac{{\pi }}{3}} \right),
\end{eqnarray}
\begin{eqnarray}
&&y_i '\left( {Q_1 ,Q_ +   = 0,Q_ -   = 0} \right) = \nonumber \\
&&\frac{{\sqrt 2 }}{{3d}}\left( {R_0  + \frac{{Q_1 }}{{3^{1/4} f}}} \right)
\sin \left( {\frac{\phi }{2} + \vartheta _i  + \frac{{\pi }}{3}} \right),
\end{eqnarray}
\begin{equation}
z_i\left( {Q_1 ,Q_ +   = 0,Q_ -   = 0} \right)=0.
\end{equation}
\end{subequations}
These equations show that the positions of the three nuclei can not
be expressed by $Q_1$, $Q_+$ and $Q_-$ when $\rho=0$, and hence the
expansion  Eq. (\ref{expandV}) is not valid as $\Delta V_e \left( {\mathcal{Q}; \mathbf{r_e}'} \right)$ is not single-valued at $Q_+=Q_-=0$ and therefore is not infinitely differentiable. However, if we define another set of
$\widetilde{x'},\widetilde{y'},\widetilde{z'}$ axis by a passive
rotation through $\phi / 2$ about $z'$-axes, the positions of nuclei
have defined values when $Q_+ = Q_- = 0$ in this new frame:
\begin{subequations}
\begin{eqnarray}
&&\widetilde{x_i '}\left( {Q_1 ,Q_ +   = 0,Q_ -   = 0} \right) = \nonumber \\
&&\frac{{\sqrt 2 }}{{3d}}\left( {R_0  + \frac{{Q_1 }}{{3^{1/4} f}}}
\right)\cos \left( {\vartheta _i  + \frac{{2\pi }}{3}} \right),
\end{eqnarray}
\begin{eqnarray}
&&\widetilde{y_i '}\left( {Q_1 ,Q_ +   = 0,Q_ -   = 0} \right) = \nonumber \\
&&\frac{{\sqrt 2 }}{{3d}}\left( {R_0  + \frac{{Q_1 }}{{3^{1/4} f}}}
\right)\sin \left( {\vartheta _i  + \frac{{2\pi }}{3}} \right),
\end{eqnarray}
\begin{equation}
\widetilde{z_i '} \left( {Q_1 ,Q_ +   = 0,Q_ -   = 0} \right) = 0.
\end{equation}
\end{subequations}
Hence, the following expansion is valid,
\begin{equation}
\widetilde K_{\Lambda \Lambda '}  = \sum\limits_\nu  {\widetilde K_{\Lambda \Lambda '}^\nu
Q_\nu  }  + \frac{1}{2}\sum\limits_{\nu \mu } {\widetilde K_{\Lambda
\Lambda '}^{\nu \mu } Q_\nu  Q_\mu  }  + \ldots,
\end{equation}
and therefore using the analysis of Longuet-Higgins \cite{Longuet}  in which $K_{\Lambda \Lambda }$ is expanded to third order in
$Q_1$ and second order in $\rho$, gives:
\begin{equation}
\widetilde K_{\Lambda \Lambda }  =
\widetilde K_{\Lambda \Lambda }^1 Q_1  + \widetilde K_{\Lambda \Lambda
}^{11} Q_1^2  + \widetilde K_{\Lambda \Lambda }^{111} Q_1^3  + \widetilde
K_{\Lambda \Lambda }^{ +  - } \rho ^2.
\end{equation}
We also have
\begin{equation}
\widetilde K_{\Lambda \Lambda  \pm 1}  = \widetilde K_{\Lambda  \pm 1\Lambda }  = 0,
\end{equation}
and
\begin{equation}
\widetilde K_{\Lambda \Lambda  \pm 2}  = \widetilde K_{\Lambda  \pm 2\Lambda
}  =  \widetilde K_{\Lambda \Lambda  \pm 2}^1 \rho e^{ \mp i\phi },
\end{equation}
to first order in $\rho$. The matrix elements of $\widetilde
K_{\Lambda \Lambda '}$ also obey
\begin{eqnarray}
\widetilde K_{\Lambda \Lambda '}  \approx  && - \pi \int {dr'{\bar {f}_{\varepsilon l \Lambda} \left({r'}\right)\bar {f}_{\varepsilon l \Lambda'}\left({r'}\right) }}\\
&& \times \int {d\widetilde \omega _e' Y_{l\Lambda }^*  \left(
{\widetilde \theta _e ',\widetilde \varphi _e '} \right)} V_e \left(
{\mathcal{Q},\mathbf{r_e}'}
\right) Y_{l\Lambda '} \left( {\widetilde \theta _e ',\widetilde \varphi _e
'} \right), \nonumber
\end{eqnarray}
where we use the tilde notation to stress that these are expressed in terms of
coordinates in the new frame. The rotation to the new frame from
our original one has the following effects: $\widetilde \theta _e  =
\theta _e$ and $\widetilde \varphi _e  = \varphi _e  - \phi /2$. Hence a comparison with Eq.(\ref{Klambda}) gives
\begin{equation}
\widetilde K_{\Lambda \Lambda '}  = K_{\Lambda \Lambda '}  e^{i\left(
{\Lambda  - \Lambda '} \right)\phi /2}.
\end{equation}
Because the reaction matrix $K$ and quantum defect matrix $\mu$ are
related by $\mu  = \mu \left( {\mathcal{Q} = 0} \right) + \arctan \left( K \right)/\pi  \approx \mu \left( {\mathcal{Q} = 0} \right) + K/\pi  + O\left( {K^3 } \right)/\pi+ \ldots$, at least to first order, the $K$ matrix and $\mu$ matrix have the same symmetry properties, and hence we write
the form of $\mu$ matrix as Eq.(\ref{mumatrix}) and Eqs.(\ref{mufit1}-\ref{mufit2}).

\end{document}